Female Librarians and Male Computer Programmers? Gender Bias in Occupational Images on Digital Media Platforms

Vivek Singh, Mary Chayko, Raj Inamdar, and Diana Floegel

School of Communication and Information, Rutgers University,

4 Huntington Street, New Brunswick, NJ, 08901.

Phone: 848-932-7500

v.singh@rutgers.edu

.

Abstract

Media platforms, technological systems, and search engines act as conduits and gatekeepers for all kinds of information. They often influence, reflect, and reinforce gender stereotypes, including those that represent occupations. This study examines the prevalence of gender stereotypes on digital media platforms and considers how human efforts to create and curate messages directly may impact these stereotypes. While gender stereotyping in social media and algorithms has received some examination in recent literature, its prevalence in different types of platforms (e.g., wiki vs. news vs. social network) and under differing conditions (e.g., degrees of human and  machine led content creation and curation) has yet to be studied. This research explores the extent to which stereotypes of certain strongly gendered professions (librarian, nurse, computer programmer, civil engineer) persist and may vary across digital platforms (Twitter, the New York Times online, Wikipedia, and Shutterstock). The results suggest that gender stereotypes are most likely to be challenged when human beings act directly to create and curate content in digital platforms, and that highly algorithmic approaches for curation showed little inclination towards breaking stereotypes. Implications for the more inclusive design and use of digital media platforms, particularly with regard to mediated occupational messaging, are discussed.

**Female Librarians and Male Computer Programmers? Gender Bias in**

**Occupational Images on Digital Media Platforms**

## Introduction

Media platforms, technological systems, and search engines are conduits and gatekeepers for all kinds of information (Bhargava and Feng, 2002; Otterbacher, Bates and Clough, 2017). They often influence, reflect, and reinforce stereotypes (Herdağdelen, 2011; Noble 2018, 2013), including those that represent occupations. Occupational choice is heavily influenced by gender; people are often steered toward and select professions on the basis of perceived gendered traits and qualities. Messages communicated and received about maleness, femaleness, technology, work, and power, including via the media, can easily become stereotypes that help to solidify these inequalities and "hierarchies of the gender order" (Wajcman 2008, p. 81). This contributes to women being underrepresented in certain occupations which tend to be better-paying and/or more respected (i.e., computer programmer, civil engineer), and overrepresented in others (i.e., librarian, nurse -- U.S. Department of Labor, 2018).

Stereotypes are "sets of socially shared beliefs about traits that are characteristic of members of a social category" (see Prot et al., 2015). They can provide a kind of shortcut for visualizing groups and categories of people and conceptualizing highly complex social constructs. When stereotypes are taken as fact and applied uncritically (as to individual members of a social category), they can lead to bias, prejudice, discrimination, and outright harm. Considerable research, however, indicates that while exposure to media-depicted stereotypes can *increase* stereotyped thinking by those who engage with them, *exposure to counter-stereotypical images can reduce stereotypical attitudes (*Prot et al., 2015; Dasgupta & Greenwald, 2001; Ramasubramanian, 2011). As the latter offers real opportunities for destabilizing gender hierarchies and contributing to social equality, there is great value in investigating the conditions under which such stereotypes prevail in the modern media landscape, and the extent to which human agency in new media use may be challenging or altering entrenched stereotypes.

While much evidence exists that depictions of occupations in mass media overwhelmingly reflect "traditional" stereotypes (Arslan & Koca, 2007; Bligh, Schlehofer, Casad, & Gaffney, 2011; Sink & Mastro 2017), the prevalence and nature of stereotypical -- *or potentially counter-stereotypical* -- images in new digital media platforms is less studied. Research by Kay, Matuszek & Munson (2015), Otterbacher, Bates & Clough (2017), and Noble (2018) indicates that gendered stereotyping via internet search engines is very much in evidence. This would seem to align with the proposition that pre-existing gendered power relations introduce inequities into technological systems so powerfully that stereotypical messages and images are bound to result. This sociotechnical perspective (Winner, 1980) recognizes that social systems and technical systems are interdependent and co-constructed, and thus suggests that inequities are introduced into the system at so many points in the process (design, content creation, curation, use) that they continually reinforce one another, becoming all but impossible to extract from the system (Emery, 1959). However, some theorize that as gendered relations develop and change over time, technological systems can and will change as well, potentially dramatically (if gradually), as they are re-designed and used by human beings in new ways that reflect and even drive new social attitudes and behavior (see Shapiro, 2015 and Wajcman, 1991 for a discussion of these theories and this debate). If the latter is the case, media stereotypes could begin to shift and change in visible, measurable ways, especially in times like the current cultural moment in which gender relations are actively being reexamined, and especially on new media platforms on which human users can directly create and curate content.

Research into bias in media and algorithms is abundant (see Noble, 2018; Kay et al., 2015; Otterbacher et al., 2017; O'Neil, 2016). This study, however, is the first research effort to explore the phenomenon of gender stereotyping with regard to a range of diverse digital platforms (Bing, Twitter, NYTimes.com, Wikipedia, Shutterstock) and occupations (librarian, nurse, computer programmer, civil engineer). It also investigates the role of human and algorithmic efforts in curating gender-biased content and looks at how these practices may (or may not) change over time.

This research, then, examines how images representing four occupations that are highly gender-segregated as per the U.S. Department of Labor statistics (2018) – two that are traditionally female (librarian and nurse) and two that are traditionally male (computer programmer and civil engineer) -- are depicted on four types of digital media platforms (Twitter, NYTimes.com, Wikipedia, and Shutterstock), from the standpoint of content creation and curation. Rather than considering human and algorithmic content creation and curation in binary fashion, this work follows a sociotechnical perspective in assuming that all modern sites require a combination of human effort and automation to populate and curate content (O'Neil, 2016). Specifically, the study explores the following research questions:

*RQ1: How different is the image-based representation of highly gender-segregated professions (librarian, nurse, computer programmer, civil engineer) in the physical world compared to digital spaces (Twitter, NYTimes.com, Wikipedia, Shutterstock)?*

*RQ2: How do the differences above vary with time?*

*RQ3: How does the relative ratio of human and algorithmic effort in content creation and curation affect the degree and types of biases observed on different types of digital media platforms?*

This research has particular relevance for the field of library and information science, especially as the shape and constitution of modern libraries and the cultural roles of librarians are concerned (Dilevko & Harris, 1997). It also advances the study of the impact of technology on diverse participation in STEM (Science, Technology, Engineering, and Mathematics) disciplines, the need for which was recently underscored by the ACLU's lawsuit against Facebook for disproportionately showing younger male users technical job ads (Tiku, 2018). Finally, the results of this study will point the way toward the development of more inclusive messaging in information created and shared on digital media platforms and other online spaces. This can enhance understanding of bias in the design and use of technological platforms and information portals, decrease the prevalence of stereotyping in mediated messaging regarding occupations and in these spaces overall, and help to bring about a more equal representation of women and men in occupations that are currently gender-segregated. As we will discuss in subsequent sections of this article, we acknowledge that our study is significantly limited because it takes a binary approach to gender, meaning we only discuss men and women in our analysis. However, we hope that our discussions of this limitation bring continued attention to wider problems with ways in which media, technical systems, and data collection instruments operate within frameworks that assume a gender binary and subsequently perpetuate inequities against nonbinary people (Keyes, 2018; Spiel, Keyes, & Barlas, 2019).

## Review of Literature

Both scholarly and popular discourses focus on (often binary) gender stereotypes perpetuated by and reflected in media. While disparities between gender representations have been reported in print media at least since the 1960s, recent research suggests that inequitable treatments of gender have only increased with the influx of digital media, despite conceptions of such platforms as liberating or emancipatory (Baily, Steeves, Burkell, & Regan, 2013; Döring, Reif, & Poeschl, 2016). This section reviews the literature on disparities across binary conceptions of gender in different digital and analog media platforms, as well as theoretical frameworks relevant to the understanding of mediated gender stereotypes.

### Gender stereotyping in newspapers and advertisements

Studies of newspaper content have found worldwide proliferation of binary gender stereotypes in articles, images, and advertisements. For example, Spanish newspapers often depict gender stereotypes in print and visual content and disproportionately feature women in shorter news items and the less prestigious Sunday news (de Cabo, Gimeno, Martínez, & López, 2014). Such depictions are linked to wider patterns of underrepresentation, stereotyping, and discrimination against women (de Cabo et al., 2014). Similarly, analysis of the Showbiz and Entertainment sections of Pakistani newspapers reveal that female subjects are discussed more than male subjects in terms of their personal lives, women are shown in more pictures than men, and women are more sensationalized overall (Rasul, 2009). Turkish daily newspapers feature images that portray women as glamorous, sexy, heterosexual, and mothers; content also favors male athletes over female athletes and content on women in sports more often includes information such as marital status than content on men in sports (Arslan & Koca, 2007). Linguistic analysis further demonstrates that adjectives and compliments used to describe women make them seem powerless and beautiful, whereas those associated with men imply strength (Rasul, 2009).

Advertisements also draw on gender stereotypes. Content analysis of advertisements for technology products sampled from professional journals in the fields of business, computing, science/engineering, and library and information science finds that men are depicted more frequently than women, although the distribution of male and female figures in various poses is more egalitarian in ads found in traditional library journals (Dilevko & Harris, 1997).

Additionally, the depictions of male and female roles in relation to technology is largely stereotypical: Men are often portrayed as deep thinkers who are connected to the future, whereas women often convey the notion of simplicity of product use (Dilevko & Harris, 1997; Döring & Pöschl, 2006).

### Gender stereotyping in social media and search engines

Studies of social media suggest that digital platforms' purported ideal to present more expansive or equitable gender constructions is often unrealized. For example, social networking sites (SNS) provide many stereotypical images of girls as sexualized objects desiring male attention (Baily et al., 2013). SNS are commoditized environments in which girls' self-exposure yields both high status and heightened judgement, fueling discriminatory standards that police girls' ability to fully participate online and engage in gender-defiant presentations (Baily et al., 2013). An analysis of sexting via Facebook and Blackberry Messenger finds that girls may experience shame or punishment for sexting while boys' social status is enhanced (Ringrose, Harvey, Gill, & Livingstone, 2013). As opposed to magazine advertisements, research suggests that selfies uploaded to SNS such as Instagram more strongly reflect gender stereotypes (Döring, Reif, & Poeschl, 2016).

Social media can, however, also spread messages, images, and expressions of support that counter gender stereotypes, including the Western binary conception of gender. Social and other digital media forms may help transgender individuals develop their identities (Craig & McInroy, 2014), given affordances that online platforms can provide for members of marginalized populations (Mehra, Merkel, & Bishop, 2004). Presence of gender role-resistant content curated by users may create informal learning environments (Fox & Ralston, 2016) or spaces for self-disclosure (Green, Bobrowica, & Ang, 2015) for queer individuals who wish to challenge gender stereotypes and the gender binary.

Search engines often also perpetuate gender stereotypes that intersect with other identity categories (e.g., race, class, disability) and associated inequities. Noble (2018) reports that, as a sociotechnical system, Google perpetuates conceptions of people and ideas in ways that strongly reflect racist and sexist stereotypes; black women, for example, are frequently associated with pornography in results generated by PageRank. Similarly, Google image search results for various careers have been reported to systematically underrepresent women and present stereotypical exaggerations of women in male-dominated professions (e.g., a sexy construction worker) (Kay et al., 2015). Further, Google has been reported to show high-income jobs to men more often than to women (Datta, Tschantz, & Datta, 2015). This is of particular concern given that algorithms are often framed as uncontestable despite their demonstrated reinforcement of discrimination, including gender discrimination (O'Neil, 2016).

### Sociotechnical systems

The ability of media to both spread and counter gender stereotypes can be contextualized in a sociotechnical frame. Conceived of in organizational studies, a sociotechnical systems approach recognizes that technology is designed by humans and will be used in a social context that shapes how it is adopted (Benders, Hoeken, Batenburg, & Schouteten, 2006; Leonardi, 2012). Thus, neither a technology nor its human designers and users fully determine how a system will function; systems are inherently fused with human biases. This perspective is closely related to that of sociomateriality, which claims that technical features and social action co-construct each other (Leonardi, 2012). For example, the socio-cultural context surrounding material affordances and constraints of social media platforms and search engines can result in a system being either more or less useful for individuals looking for information about their gender identities and sexualities (Kitzie, 2017). However, while sociomateriality focuses on platforms' technical features (e.g., a search box), a sociotechnical systems view presents a more holistic perspective on systems and how they function.

The fusion between humans and technical systems also influences how individuals experience various media. The data generated in the use and consumption of digital media provide a unique window on the world as it is experienced, shaped, and comes to be understood by millions of users (see Chayko, 2018; Baym, 2015). All forms of media, as agents of socialization, provide individuals with the possibility to learn social norms and values related to both broad social categories such as gender, race, and occupation, as well as more microinteractional phenomena such as making day-to-day decisions regarding practices, beliefs, and relationships (Genner and Suss, 2017; Prot et al., 2015). In the process, people's attitudes and behaviors are constructed, shared, and then continue to affect a system in a cyclical process, thus contributing to a sociotechnical co-construction of humans and technology. These attitudes and behaviors are often influenced by one's gender, the meaning and nature of which is in itself also continuously being shaped (see Butler, 1988; Shapiro, 2015).

## Methodological Approach

This study compared the number of images of men and women associated with four occupations (librarian, nurse, computer programmer, civil engineer) on four media companies' digital platforms (Twitter, NYTimes.com, Wikipedia, Shutterstock) to the rates of gender representation in these occupations according to national labor statistics (U.S. Department of Labor, 2017). All data were collected using Microsoft Bing search engines Application Programming Interface (API) which allows for getting search results from specific sites (e.g. NYTimes.com); the results from Bing without the specification of any sites were used for baseline comparison.

 A determination was then made as to whether the results break from or replicate national trends regarding bias and stereotyping in binary gender-segregated occupations. These large, influential, yet different kinds of digital sites were selected because they capture images and text for different and distinct purposes: one provides a platform for social media networking (Twitter), one presents the news along with editorial and feature stories (NYTimes.com), one is a crowdsourced encyclopedia (Wikipedia), and one provides consciously framed photographs for a variety of uses (Shutterstock).

While Facebook and Google are obviously highly influential media companies and platforms, the investigators elected not to use them as research sites for this study, selecting Twitter and Bing instead. Researchers studying Facebook (and its photo-sharing app Instagram) are subject to growing and frequently shifting restrictions which are much more restrictive than Twitter (Bastos & Walker, 2018). And researchers working on issues similar to those explored here have reported that Bing provides much more robust user-friendly API support than does Google (Otterbacher et al., 2017). The investigators expect, however, that most, if not all, of the results of these studies will have broad applicability across media platforms and sites, including Facebook and Google.

We use the terms "men" and "women" in this paper as these are the terms used by Clarifai, the API used to obtain gender labels (see the "Implementation" section). The U.S. Department of Labor's data collection instruments exclude nonbinary individuals as well, recognizing the same two gender categories. We acknowledge the ways in which this impacts our results: 1) the images Clarifai classifies may not depict men or women, given that gender presentation does not always correlate with gender identity, and Clarifai uses recognition software based on normative conceptions of what "men" and "women" look like; 2) this work perpetuates binary conceptions of gender that exclude nonbinary individuals and are harmful for them and for society writ large. Both of these limitations are expanded upon in the "Discussion" section of this paper. Despite these limitations, though, we believe that this work advances the understandings of gender stereotypes in media spaces while drawing needed attention to problematic epistemic values embedded in systems that perpetuate gender inequities and attempt to quantify gender (Keyes, 2019).

In this work, we pay specific attention to the process of content creation and content curation across platforms. Content creation across all platforms requires humans to contribute content. Twitter is considered most inclusive of amateur content, undertaking minimal quality checks on content (Naaman, Boase, & Lai, 2010). While some content providers on Wikipedia may be amateurs, they do need to meet certain thresholds of quality, as opposed to contributors to Twitter (Stvilia, Twidale, Smith, & Gasser, 2008). Content on NYTimes.com and Shutterstock, on the other hand, is specifically created by specialists and vetted for quality and "organizational voice." This continuum is summarized in Figure 1.

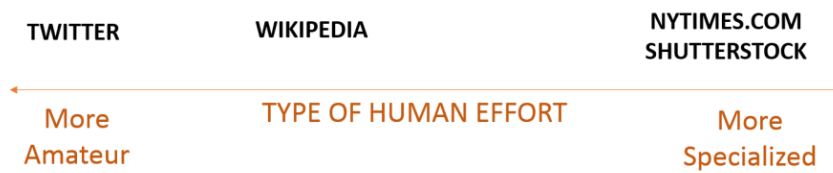

*Figure 1.* Relative position of the different digital platforms on the continuum for the type of human effort (amateur to specialized) in content creation.

Consistent with the sociotechnical perspective, we suggest that the curation process of content on these platforms requires interlocking human and algorithmic contributions. Figure 2 indicates the relative position of the different digital platforms on the human effort – algorithmic effort continuum for content creation and curation. Given its massive scale of elements (tweets) to process and real-time primacy model, Twitter curation process is largely automated; while humans constructed, maintain, and continue to shape Twitter's algorithms, they are not moderating content as they do for platforms like Wikipedia. An algorithm based on reverse chronological ordering was Twitter's default presentation mode for a long period of time and temporal recency is still one of the important factors in its content curation process (Binder, 2018). However, content is vetted quite extensively by human moderators on Wikipedia (Stvilia, Twidale, Smith, & Gasser, 2008) and presumably on professional media companies like NYTimes.com and Shutterstock. This study considers Twitter to involve mostly algorithmic effort in content curation, while Wikipedia, NYTimes.com, and Shutterstock are more likely to require more direct significant human effort in content curation. This does not mean that NYTimes.com does not apply algorithms in curation but simply indicates that NYTimes.com articles typically require more direct human effort in curation than those found on Twitter (see Figure 2).

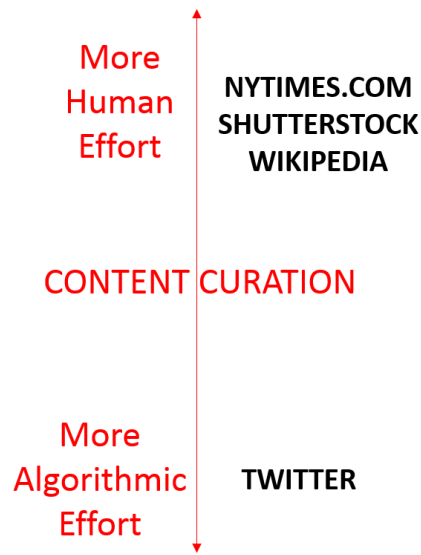

*Figure 2.* Relative position of the digital platforms on the human effort – algorithmic effort continuum for content curation.

**Implementation**

For the analysis in this work, the top 100 relevant images were downloaded for each profession from each platform using Bing API, then passed through the Clarifai API to obtain gender labels. Potential errors in gender labels were then quantified and the difference between expected (e.g., based on labor statistics or an equal ratio) and observed binary gender ratios was computed. Details follow below on the data collection, filtering, gender label assignment, quantifying the potential errors, and baseline comparison.

All images were downloaded from Bing image search using the publicly available Bing image search API. Collecting images directly using the search APIs or interfaces for the various platforms (Twitter, NYTimes.com etc.) was also considered, but only Twitter and Shutterstock allowed for a specific search of *images* pertaining to the keywords, while NYTimes.com and Wikipedia allowed only for broader searches within their systems, not for searches focusing only on images. Hence, for consistency, we decided to search for images using a common Bing API where the platform considered was indicated as a search parameter as described below.

To download images from Bing, both query parameters and image filters were needed.

*Query:*

The query search term on Bing image search is of the format – "<Job> site:<domain>".

Example- "Librarian site:twitter.com" downloaded the images pertaining to keyword librarian from twitter.com.

For baseline Bing search images, sites are not mentioned. Thus, the query was simply: "Librarian".

*Image Filters:*

We use filtering options- "imageType": "Photo", "imageContent": "Portrait", and "Safesearch": "Moderate". Image results on Bing image search may consist of cliparts, animations or logos in addition to photographs. These kinds of images were not included in the results due to the above parameters. Adult content images were also not found in the results presumably because of the "safe search" option. Filters focused on images which are photographs and in portrait format. All images available (up to 100) upon searching the query on Bing image search after applying the abovementioned filters were downloaded for each job across all the platforms under consideration. Note that we do not have control over the process used by Bing or the underlying websites (e.g., Twitter.com) to provide images in the search results. For instance, the results for the keyword "Librarian" could be based on the text in the tweets, a computer vision algorithm that searches for librarian like appearances, or something else. Hence, interpreting individual results is difficult; we focus on comparisons across the results coming from different websites for which the collection and analysis process has been kept consistent.

Clarifai demographics API assigns a subject a male or female label in a given image based on its computer vision algorithm (Clarifai, 2018). If the image has multiple subjects, only the subject that is closest to the center of the image was considered. The use of the visual imagery-based API allows only external, visible markers of identities to be captured; it considers these to be proxies for gender (see Methodological Approach," as well as the "Discussion" and section, for why this is problematic). Gendered visual markers, however flawed, are used by both experts and novices (including children) for making sense of the world around them (Goffman, 1978; Singer & Singer, 2012) because social institutions such as schools and governing bodies often instantiate binary conceptions of gender and normative assumptions about gender presentation (Gowen & Winges-Yanez, 2014). The Clarifai API also provides a confidence score which states how accurately it rates its own binary gender assignment. For this study, only those images with subjects Clarifai identified as men or women with a confidence score greater than 90 were considered. While these settings reduced the potential errors within Clarifai's own ratings of its gender assignments, some did remain. The next section (specifically, subsection labeled "Quantifying Errors") presents the approach adopted to interpret errors in the results and estimate the error bounds. Though we use "men," "women," "male," and "female" throughout the results section, we acknowledge that these labels are assigned to a subject by the Clarifai API and may not be accurate to the individuals in images.

This study interprets the results based on two metrics discussed in the recent literature on algorithmic bias (Calmon, Wei, Vinzamuri, Ramamurthy & Varshney, 2017). *Demographic parity* suggests that the representation of the different demographic groups (here, men and women) should be equal. *Equal opportunity* suggests that the number of images for each gender successfully selected by the algorithm should be in proportion to the number of candidates from each class that are eligible (this study considers *labor statistics* to be such a ratio). Hence, the female percentages for each job across all social media sites were evaluated in two ways:

1) Comparing results with the labor statistics
2) Comparing results with a scenario including equal representation for the considered genders

All the data considered in the study were collected and analyzed twice in this work - once in the summer of 2018 and once in the summer of 2019. Search data in this study were compared with the last available labor statistics (i.e., 2017 and 2018) data provided by the US Bureau of Labor Statistics for each of the considered professions before the search data were collected (U.S. Department of Labor, 2018). Collection of data across the two years allows for higher confidence in the obtained results and also allows for the analysis of trends across time.

The results are as shown in Table 1 A, B, C. The results indicate the percentage of image results, which were considered to represent women in the search results for the four professions and the four digital platforms. For baseline comparison purposes, we also report the results for labor statistics, Bing platform (without any site specified for filtering), and the search results for the terms "men" and "women."

Table 1A

Comparison of results between labor statistics and considered digital platforms for search results.
(Average values across years 2018 and 2019)

| Job | Labor Statistics | Bing (without site filters) | Twitter | NYTimes | Wikipedia | Shutterstock |
|---|---|---|---|---|---|---|
| **Librarian** | 79.0 | 77.0 | 69.5 | 45.5 | 24.5 | 74.0 |
| **Nurse** | 89.3 | 90.5 | 83.0 | 57.5 | 71.0 | 98.5 |
| **Computer Programmer** | 21.2 | 39.5 | 13.0 | 14.9 | 9.5 | 34.0 |
| **Civil Engineer** | 14.6 | 14.5 | 14.5 | 39.1 | 2.5 | 27.0 |
| **Woman** | 100.0 | 99.5 | 99.0 | 97.0 | 91.0 | 99.5 |
| **Man** | 0.0 | 0.5 | 1.5 | 2.0 | 3.5 | 0.0 |

Table 1B

Comparison of results between labor statistics and considered digital platforms for search results in Summer 2018

| Job | Labor Statistics | Bing (without site filters) | Twitter | NYTimes | Wikipedia | Shutterstock |
|---|---|---|---|---|---|---|
| **Librarian** | 79.5 | 74.0 | 71.0 | 50.0 | 22.0 | 74.0 |
| **Nurse** | 89.9 | 87.0 | 83.0 | 51.0 | 71.0 | 100.0 |
| **Computer Programmer** | 21.2 | 31.0 | 8.0 | 19.2 | 9.0 | 34.0 |
| **Civil Engineer** | 14.4 | 13.0 | 10.0 | 47.6 | 2.0 | 28.0 |
| **Woman** | 100.0 | 100.0 | 98.0 | 98.0 | 92.0 | 99.0 |
| **Man** | 0.0 | 1.0 | 1.0 | 3.0 | 2.0 | 0.0 |

Table 1C

Comparison of results between labor statistics and considered digital platforms for search results in Summer 2019

| Job | Labor Statistics | Bing (without site filters) | Twitter | NYTimes | Wikipedia | Shutterstock |
|---|---|---|---|---|---|---|
| **Librarian** | 78.5 | 80.0 | 68.0 | 41.0 | 27.0 | 74.0 |
| **Nurse** | 88.6 | 94.0 | 83.0 | 64.0 | 71.0 | 97.0 |
| **Computer Programmer** | 21.2 | 48.0 | 18.0 | 10.6 | 10.0 | 34.0 |
| **Civil Engineer** | 14.8 | 16.0 | 19.0 | 30.6 | 3.0 | 26.0 |
| **Woman** | 100.0 | 99.0 | 100.0 | 96.0 | 90.0 | 100.0 |
| **Man** | 0.0 | 0.0 | 2.0 | 1.0 | 5.0 | 0.0 |

All figures represent <u>woman percentage</u>. Thus, as per Table 1A, 77% of the "Librarian" search results on Bing (without any site filters), were judged to depict women.

### Quantifying Errors

As a baseline to quantify the errors in the process "Man" and "Woman" were searched in each of the considered settings. Results were expected to be, respectively, 100% male subjects and 100% female subjects for these queries. The results, however, were not always exactly 100%. For instance, as shown in Table 1A, 99% female subjects were obtained after searching for "Woman" on Twitter, indicating 1% fewer female subjects than expected. Similarly, a search for "Man" on Twitter retrieved 1.5% female subjects, indicating 1.5% more female subjects than expected. Such errors are typically less than 5% (average = 1.9%), which provides a reasonable amount of confidence in the analysis process. The errors obtained for "Male" and "Female" are considered the error margins and strong claims are avoided within those margins of errors for any subsequent analysis.

"Librarian" obtained 69.5% female subjects on Twitter, which is 9.5% less than labor statistics for women in librarianship. Because 1% fewer female subjects were retrieved when querying for "Woman" and 1.5% more female subjects were retrieved when querying for "Man" during the librarian search on Twitter, the observed 9.5% difference between Twitter images and labor statistics should be in the range from 8.5% to 11%.

Note that the stereotypes and biases found in this study can be attributed to multiple factors. For instance, the discrepancies in the NYTimes.com results could be a function of the content creation process, the curation process, the Bing processing of NYTimes.com image results, and the Clarifai API for identifying the men and women in the pictures. As shown in Table 1, in most cases the Bing results (without site filters) were closer to the labor statistics than those obtained using Bing with specific filters (e.g. Twitter.com). This suggests that specifying site filters led to more pronounced disparities, and a comparison of those disparities across sites can be indicative of the kind of content present on those sites. Note that because the cause of each error cannot be determined, focus remains on the relative analysis across different platforms wherein the data collection and analysis process has been kept consistent.

**Comparing results with labor statistics**

To evaluate performance of different digital platforms with respect to labor statistics, the media platform result was subtracted from the labor statistics (averaged over the years 2018 and 2019) as shown in Table 2. For instance, the Twitter performance for "librarian" compared to actual labor statistics would be 69.5-79.0 = -9.5.

Table 2

Differences between the ratio of gender representation on platforms as compared with actual labor statistics

| Job | Labor Statistics | Twitter | NYTimes | Wikipedia | Shutterstock |
|---|---|---|---|---|---|
| **Librarian** | 79.0 | -9.5 | -33.5 | -54.5 | -5.0 |
| **Nurse** | 89.3 | -6.3 | -31.8 | -18.3 | 9.3 |
| **Computer Programmer** | 21.2 | -8.2 | -6.3 | -11.7 | 12.8 |
| **Civil Engineer** | 14.6 | -0.1 | 24.5 | -12.1 | 12.4 |
| **Woman** | 100.0 | -1.0 | -3.0 | -9.0 | -0.5 |
| **Man** | 0.0 | 1.5 | 2.0 | 3.5 | 0.0 |
| Average for top 4 rows | 51.0 | -6.0 | -11.8 | -24.1 | 7.4 |

The following trends are noted with regard to these results:

**Direction of errors**

1) Always fewer women
   - Twitter and Wikipedia always show fewer female subjects for all occupations
2) Generally fewer women
   - NYTimes.com generally shows fewer female subjects than actual labor statistics except for civil engineer
3) Generally more women
   - Shutterstock generally shows more female subjects than actual labor statistics except for librarian

**Magnitude of errors (considering the direction)**

1) No to little error (within 5%)

- N/A
2) Moderate errors (5% to 20%)
    - NYTimes.com and Twitter in general underrepresented females moderately
    - Shutterstock overrepresented females moderately
3) Significant errors (20% or higher)
    - Wikipedia results show significant divergence (underrepresentation) from the labor statistics

**Comparing results with gender representation**

In this dataset, two of the professions (librarian and nurse) are heavily occupied by women and the other two (civil engineer and computer programmer) by men.

Hence when quantifying the magnitude of errors, the gendered nature of the jobs and whether the errors observed are breaking or reinforcing stereotypes can be considered. For instance, it can be argued that NYTimes.com underrepresents women in nursing-related photos because it wants to provide more equal gender representation in its content. The same could be argued regarding the overrepresentation of women as civil engineers.

Therefore, in the Table 3 below, positive credit is given for errors if they are in the direction of 50:50 parity; they are scored negatively otherwise.

Table 3
Comparison of results with actual labor statistics (giving credit for the challenging of stereotypes)

| Job | Labor Statistics | Twitter | NYTimes | Wikipedia | Shutterstock |
|---|---|---|---|---|---|
| **Librarian** | 79.0 | 9.5 | 33.5 | 54.5 | 5.0 |
| **Nurse** | 89.3 | 6.3 | 31.8 | 18.3 | -9.3 |
| **Computer Programmer** | 21.2 | -8.2 | -6.3 | -11.7 | 12.8 |
| **Civil Engineer** | 14.6 | -0.1 | 24.5 | -12.1 | 12.4 |
| Average for top 4 rows | | 1.9 | 20.9 | 12.2 | 5.2 |

The following trends were noted:

**Magnitude of errors (giving credit to challenging stereotypes)**

1) Breaking stereotypes towards equal representation (scoring above 20 points)

- NYTimes.com

2) Challenging stereotypes towards equal representation (between 5 and 20 points)

- Wikipedia and Shutterstock

3) Little to no movement towards equal representation (less than 5 points)

- Twitter

## Discussion

The results of this study suggest that some gender-based occupational stereotypes are being challenged on some digital media platforms (e.g. NYTimes.com as shown in Table 3). Other gender stereotypes and biases, however, are being reinforced. Here we revisit the three research questions and discuss the results obtained.

*RQ1: How different is the image-based representation of highly gender-segregated professions (librarian, nurse, computer programmer, civil engineer) in the physical world compared to digital spaces (Twitter, NYTimes.com, Wikipedia, Shutterstock)?*

Consistent with many previous studies, this study finds that women (or, at least, subjects the API assigns as women) are largely underrepresented in images on digital platforms. There were clear exceptions, as discussed below, but the general trend holds. Remarkably, this underrepresentation was consistent for both male-dominated and female-dominated professions.

Examining the data based on whether creators are amateurs or specialized provides an interesting perspective (Figure 2). With regard to error directions in the representation numbers and labor statistics (Table 2), platforms with largely amateur contributors (Twitter, Wikipedia) underrepresent women across all professions. This may relate to these systems' sociotechnical contexts. A dearth of female editors on Wikipedia, for example, may lead to less frequent and more stereotypically charged coverage of women across its content (Hill & Shaw, 2013). Most recently, among three Nobel Prize-winning scientists, only the woman scientist lacked a Wikipedia page until the awarding of her prize; pages for the men had already been created (Nechamkin, 2018). Though subsets of Twitter users may certainly engage in equity-oriented work, a study of discourse among a large sample of users found that male-female power inequities, including "mansplaining" and other gender-specific language, are prevalent on the platform (Bridges, 2017). In the aggregate, results from these platforms suggests that gender stereotypes are far from eradicated on digital media, just as they are far from absent in societal discourse.

Regarding the magnitude of these errors (see Table 2), Twitter's divergence from actual labor statistics – the gendered composition of those occupations in reality -- is lowest. Images on these platforms correlate most precisely to the demographic makeup of those occupying the jobs examined here. On Twitter, as on Wikipedia, more images of men are found representing all occupational categories (though men and women seem to use Twitter in roughly equal numbers – see Smith and Anderson, 2018). The difference between the two platforms arguably lies in the terms of the level of automation in curation. While Twitter largely sorts the results based on recency without any major filtering, Wikipedia content is curated at length by human content moderators and curators. This aspect is discussed further under RQ3.

*RQ2: How do the differences above vary with time?*

RQ2 of this work refers to the changes observed over time. As can be seen in Tables 1B and 1C, the trends have remained largely consistent over the yearlong gap in which the data for this work were analyzed. Firstly, the labor statistics-based participation across men and women remained consistent with an average change of less than 1%. Further, the abovementioned trends in terms of direction of errors, magnitudes of errors (considering the direction), and magnitude of errors (giving credit to breaking stereotypes) remained largely consistent (see tables 1B and 1C). An overwhelming majority of the platforms stayed in the same bins across the two years, and the overrepresentation and underrepresentation across professions was also quite similar. There were two exceptions. First, Twitter moved from "Always fewer females" in 2018 to "Generally fewer females" in 2019 in terms of direction of errors. It also moved from "Moderate errors (5% to 20%)" in 2018 to "No to little error (within 5%)" in 2019 in terms of the magnitude of errors (considering the direction). Second, NYTimes.com moved from "Breaking stereotypes towards equal representation (scoring above 20 points)" in 2018 to "Challenging stereotypes towards equal representation (between 5 and 20 points)" in 2019 in terms of magnitude of errors (giving credit to challenging stereotypes). While the changes in Twitter over the year are encouraging, the changes in NYTimes.com are less so. Nevertheless, NYTimes scores the highest in terms of challenging the stereotypes in both years. The results paint a cautiously optimistic horizon for the representation of women across platform. While there is some movement in the positive direction, the trends overall remain consistent. We hope that multiple works help shine more light on these disparities and motivate transformative changes by system designers across platforms in the near future to yield larger equality across genders in results.

*RQ3: How does the relative ratio of human and algorithmic effort in content creation and curation affect the degree and types of biases observed on different types of digital media platforms?*

As indicated in Table 3, NYTimes.com, Shutterstock, and Wikipedia challenged binary gender stereotypes most successfully. On the other hand, Twitter replicated, and, in some cases, reinforced binary gender stereotypes and biases.

Images on NYTimes.com diverge from the labor statistics most substantially, providing images of civil engineers who are women, librarians who are men, and nurses who are men to a much greater extent than reflected by the Bureau of Labor Statistics (although this trend does not hold for images of women as computer programmers). Shutterstock seems to be making a more modest effort toward presenting gender representations that depart from the actual labor statistics, with the exception of an insufficient number of men being represented as nurses. The magnitude of gender representations that challenge labor statistics is also modest on Wikipedia, although women subjects are underrepresented in all jobs on the platforms, including nurses and librarians.

As posited in RQ3, we interpret the results using the axes of *content creation* and *content curation* (see Figures 1 and 2). We find that the magnitude of errors compared to *actual labor statistics* showed an increasing trend based on the degree of human involvement in the curation (i.e. from mostly algorithmic curation to mostly non-algorithmic human curation). At the same time, the magnitude of error (while giving credit to the challenging of stereotypes) showed the opposite trend. *While highly algorithmic approaches for curation showed little inclination towards challenging stereotypes, those with largely non-algorithmic human curation process were more likely to do so.*

NYTimes.com and Shutterstock, as companies that deal with media and images on a very large scale (though with very different purposes), tend to speak with somewhat more unified editorial "voices" than does Wikipedia (and more so than Twitter). This may result in the somewhat more careful, less stereotypical curation of the images than was seen on the other platforms examined.

As is the case with the NYTimes.com and Shutterstock, Wikipedia is curated and vetted for content, but its entries are submitted by a wide variety of people who act independently. There is no expectation of a unified editorial vision throughout Wikipedia; oversight focuses on accuracy and sourcing of independently created entries (Wikipedia, 2018). Thus, as was found here, a wider range of types of images to accompany entries, including both stereotypical images and those that defy stereotypes, is expected. More images of men than women accompany all occupational categories on Wikipedia, which exacerbates the bias against women as civil engineers and computer programmers, while bolstering the expectation that women will work as nurses and librarians. This may be influenced by the practice of Wikipedia being primarily written by, and edited by, men (over 85% of Wikipedia contributors are men by one oft-quoted estimate from Lih, 2015).

While gender stereotyping certainly persists on digital media platforms, there are sites and conditions under which these stereotypes are beginning to be challenged (e.g. expert human curation on NYTimes). Given that exposure to counter-stereotypical images can reduce stereotypical attitudes, this enhances opportunities for gender-related social change.

**Implications**

These findings have import and resonance. First, any indication that gendered occupational stereotypes are being countered and challenged in the new media environment is striking. The reinforcement of gender stereotypes hinders progress in desegregating occupations and can discourage people from striving for careers that they may otherwise be well suited for because of their gender, especially as mediated platforms serve as information resources (Chatman, 1996). More diverse participation in STEM disciplines and in library and information-related fields would be a welcome result of this research.

Activists and others continue to challenge inequities that fuel stereotypes on both institutional and more individual levels. For example, though more popular iterations of "Me Too" have been critiqued for succumbing to individualist perspectives and commodification (Banet-Weiser, 2018), the original movement founded by Tarana Burke aimed to accelerate people's awareness and understanding of gender-based sexual harassment, abuse, and violence (Burke, 2018). Trans Lifeline, a national trans-led organization headed by Elena Rose Vera at the time of this article's publication, provides service, support, advocacy, and education for trans individuals through justice-oriented and collective community aid (Trans Lifeline, 2019). It is possible that editors and specialized creators and curators of digital media content (and their advertisers) may be becoming sensitized to such efforts and their goals. They may be looking to some extent to drive substantive change, deciding, strategically, to be more welcoming to and inclusive of individuals other than cisgender men given the present cultural climate (Roderick, 2017). That may explain some of the countering of stereotypes found in this study. Of course, some media platforms may also sincerely wish to be less overtly biased and to portray occupations in a more gender-inclusive way. However, it is imperative to note that individuals—and especially members of queer populations—continue to experience heightened discrimination and harassment online (Powell, Scott, & Henry, 2018).

Some individuals may also be becoming more sensitized to their own gender biases given recent attention to this topic. In alignment with the body of sociotechnical work in which technology and humans co-construct one another, this could explain some of the less biased outcomes seen here. The biases in these platforms are a reflection of their human designers, the technological systems they are a part of, and the users of these systems. If and as these biases change over time, systems and platforms may begin to be designed in more gender-inclusive ways over time, and the next generations of digital platform users may observe far fewer stereotypical images. To the extent that human beings can create and curate their own content, these biases can be reflected differently and even change over time. However, addressing inequities in technical systems is no simple task given that scientific and technological development are infused with longstanding normative assumptions about gender (Hoffman, 2017).

This work also contributes to the ongoing conversation on bias in algorithms. While multiple scholars have argued that algorithms and automation significantly amplify inequities (e.g., Day, 2016; O'Neil, 2016; Noble, 2018), there has also been a recent counterargument on benchmarking such algorithmic bias with human biases (Miller, 2018). Rather than arguing a case for the benefits or hazards of algorithms, this work hopes to add nuance to this conversation. Biases occur in multiple forms and vary across media platforms. Similarly, human effort in the curation process is also a continuum. This research points out the value of considering the level and type of human-decision making in the curation process. While more automation appears to replicate societal conditions, more careful non-algorithmic human curation might help *challenge* stereotypes towards gender parity. Different types of curation may surely be more appropriate for different digital platforms, companies, or users, depending on their goals.

### Limitations and Future Work

Though this work discusses biases and stereotypes, both our results and their limitations indicate wider algorithmic inequities beyond highly individualized conceptions of "bias," "stereotypes," and "fairness" (Dave, 2019). While the overrepresentation of images that align with normative conceptions of masculine-presenting bodies is certainly an important finding, the process through which we reached that conclusion reveals wider systemic harms perpetuated by technical systems that cannot account for gender as a spectrum (Keyes, 2018; 2019). Both Clarifai's API and the US Department of Labor statistics operationalize gender in binary terms that exclude nonbinary individuals, meaning that analysis derived from their data perpetuates significant epistemic harms (e.g., erasure, misrepresentation, perpetual normative conceptions of both gender and gender presentation). In essence, programs like the Clarifai API act like institutions that assign gender based on physical features that do not, in actuality, dictate a person's gender. Thus, recognition software like Clarifai reproduces normative, binary conceptions of gender (Keyes, 2018). It also parses images without the consent of those photographed, and thus has implications for consent, privacy, and ongoing problems and inequities related to facial recognition software and ways in which it may be used to police marginalized populations, including trans and nonbinary people (Gates, 2015).

Though our present study's results do not challenge this systemic epistemic violence, we hope to draw continued attention to these limitations and promote scholarship that brings a more inclusive orientation to work in human-computer interaction and information science, in order to destabilize assumptions about gender and technology and work toward equity and justice (see Spiel et al., 2019). We call for future work that aligns with these perspectives and addresses inequities beyond more simplistic conceptions of bias, fairness, and binaires (Davis, 2019; Hoffman, 2019).

We also acknowledge that this focus on gender does not embrace intersecting identity categories—such as race, class, sexuality, age, and disability—that complicate equitable representation across various professions and media platforms (Crenshaw, 1991). Librarianship, for example, has a significant underrepresentation of people of color (Kim & Sin, 2008). Future work, including that of the authors, can and will include a more intersectional perspective to understand a fuller range of occupation-based inequalities and stereotypes in the media.

Despite these limitations, this work moves the literature forward in multiple ways. It is the first concerted effort at studying how different types of platforms, and content creation and curation efforts, reinforce or challenge mediated gender biases. It also provides new and important insights into the role of human content creation and curation in a highly algorithmic – and highly gender-segregated -- digital environment. Further research that would reveal, explain, and advance understanding of the design and use of sociotechnical systems so as to highlight issues of social inclusion and justice continues to be much needed.

**Conclusion**

As human beings continue to produce and consume digital information online, the gendered imagery found in many of these messages can shape and influence human attitudes, perceptions, behaviors, and norms. This study is one of the first to examine gender variations in occupational imagery across digital media platforms and information portals. Results indicate that stereotypes are most likely to be challenged when human beings act directly to create and curate content in digital spaces that are less controlled by algorithms. At the same time, highly algorithmic approaches reinforce societal realities more so than did largely human efforts, although the two cannot be fully separated. This work adds nuance to the discussion of the biases found in content curation and identifies the scenarios in which certain gradations of algorithm-heavy or human-heavy curation may be better suited. And it sheds light on ways to improve the creation and curation of digital content such that existing stereotypes can be countered in favor of gender representations that are more equitable, paving the way for a more equitable and just society.